\documentclass[aps,pra,twocolumn,floatfix,superscriptaddress,longbibliography]{revtex4-1}
\usepackage{amsmath}
\usepackage{physics}
\usepackage[dvipsnames]{xcolor}
\usepackage{amsmath,amssymb}
\usepackage{mathtools}
\usepackage{hyperref}
\usepackage{newfloat}
\usepackage[super]{nth}
\usepackage{placeins}

\usepackage{amsfonts, amsmath, amssymb}
\usepackage[utf8]{inputenc}
\usepackage[english]{babel}
\usepackage{graphicx}
\usepackage{dcolumn}
\usepackage{hyperref}
\usepackage{bm}
\usepackage[caption=false]{subfig}
\usepackage{color}
\usepackage{soul}
\usepackage{mathtools}
\usepackage{float}
\usepackage{comment}

\newcommand{\mL}[0]{\mathcal{L}}

\newcommand{\ham}[0]{\hat{H}}

\DeclareMathOperator{\Var}{\mathbb{V}ar}
\DeclareMathOperator{\Cov}{\mathbb{C}ov}
\DeclareMathOperator{\orho}{\hat{\rho}}

\newcommand{\oa}[0]{\hat{a}}
\newcommand{\oad}[0]{\hat{a}^{\dagger}}

\usepackage{braket}

\begin{document}
\title{Optimal stochastic unraveling of disordered open quantum systems: \\ 
application to driven-dissipative photonic lattices}

\author{Filippo Vicentini}
\affiliation{Laboratoire Mat\'{e}riaux et Ph\'{e}nom\`{e}nes Quantiques, Universit\'{e} Paris Diderot, CNRS-UMR7162, 75013 Paris, France}

\author{Fabrizio Minganti}
\affiliation{Laboratoire Mat\'{e}riaux et Ph\'{e}nom\`{e}nes Quantiques, Universit\'{e} Paris Diderot, CNRS-UMR7162, 75013 Paris, France}

\author{Alberto Biella}
\affiliation{Laboratoire Mat\'{e}riaux et Ph\'{e}nom\`{e}nes Quantiques, Universit\'{e} Paris Diderot, CNRS-UMR7162, 75013 Paris, France}

\author{Giuliano Orso}
\affiliation{Laboratoire Mat\'{e}riaux et Ph\'{e}nom\`{e}nes Quantiques, Universit\'{e} Paris Diderot, CNRS-UMR7162, 75013 Paris, France}

\author{Cristiano Ciuti}
\affiliation{Laboratoire Mat\'{e}riaux et Ph\'{e}nom\`{e}nes Quantiques, Universit\'{e} Paris Diderot, CNRS-UMR7162, 75013 Paris, France}

\date{\today}

\begin{abstract}
We propose an efficient numerical method to compute configuration averages of observables in disordered open quantum systems whose dynamics can be unraveled via
stochastic trajectories.  We prove that the optimal sampling of trajectories and disorder configurations is simply achieved by considering one random disorder configuration for each individual trajectory. As a first application, we exploit the present method to the study the role of disorder on the physics of the driven-dissipative Bose-Hubbard model in two different regimes: (i) for strong interactions, we explore the dissipative physics of fermionized bosons in disordered one-dimensional chains; (ii) for weak interactions, we investigate the role of on-site inhomogeneities on a first-order dissipative phase transition in a two-dimensional square lattice.  \end{abstract}

\maketitle

\section{Introduction}
\label{sec:introduction}

The simulation of open quantum many-body systems is a formidable task.
This is due to the exponential growth of the Hilbert space with the system size combined with the need of representing mixed states \cite{BreuerBook07}.
While systems in contact with the surrounding environment are expected to relax towards a (generalized) Gibbs ensemble \cite{Ilievski2015,Langen2015}, in the most general scenario a system may not thermalize and evolve towards a non-equilibrium state.
This situation can be reached by driving the system away from its thermodynamic equilibrium \cite{Lee13,LeBoitePRL13,CarusottoRMP13} or by properly engineering the system-bath couplings in order to let the system evolve toward some target attractor of the dynamics \cite{Kapit2017,Muller2012,Diehl2008,MorigiEPL2011}.    

In the last decade, impressive experimental advances made possible to study the open nonequilibrium dynamics of large ensambles of quantum particles.
There exist several experimental platforms that allow to explore this physics, such as photonic cavities \cite{JacqminPRL2014,RodriguezPRL2017,CarusottoPRB2005}, superconducting resonators \cite{HouckNP2012,Fitzpatrick17,LabouviePRL2016}, optomechanical resonators \cite{AspelmeyerRMP14} and Rydberg atoms \cite{LeePRL13,ViteauPRL12,SaffmanRMP10}. 
Their high versatility allows to simulate the physics of several interesting Hamiltonians \cite{GeorgescuRMP86}, where most of the dynamical, spectral and correlation properties would be otherwise difficult to access. 
Indeed, their lossy nature has been repeatedly exploited in order to directly probe their state by observing the emission in the environment. 
For this reason, the stabilization and control of non-trivial, nonequilibrium many-body states in open systems has attracted an increasing number of theoretical works (see e.g. \cite{Jin2018_ising,RotaArx2018,Ma2018,Biella2017,Lebreuilly2017,Schiro2016,Jin16,WilsonPRA16,Bardyn2013,Rancaglia2010} and references therein).
Upon the variation of an Hamiltonian parameter and/or of the coupling rate with the surrounding environment, these systems can be driven across a (dissipative) phase transition in a properly defined thermodynamic limit \cite{Minganti2018}.
The peculiar behavior at criticality and the mechanism triggering the transition do not always follow the paradigms of equilibrium statistical mechanics \cite{Rota17,Marino2016,ScarlatellaArxiv2018,MorigiPRL2016}. 

This scenario can be further enriched if one considers the effects of disorder.
For closed systems, the interplay of interaction and disorder in a quantum many-body system can lead to remarkable effects such as the breakdown of ergodicity due to quantum effects, a phenomenon known as many-body localization \cite{Basko2006,Oganesyan2007,Nandkishore2015}.
Similar studies have recently been performed in the context of open systems \cite{Xu2018,Vakulchyk2018,Maximo2015,Celardo2017,Biella2013,Celardo2012,SeeArxiv18} in order to unveil the competition between dissipation and localization phenomena.
However, the theoretical investigations of the nature of far-from-equilibrium correlated phases in the presence of disorder have only scratched the surface \cite{Maghrebi2017}.
Moreover, while the non-equilibrium extension of stochastic mean-field theory has been developed \cite{Kulaitis2013}, the scarcity of effective numerical and analytical tools to tackle the problem beyond the mean-field paradigm has limited the study of the effects of disorder on extended systems. 
Indeed, dealing with disordered systems introduces an additional layer of complexity with respect to the clean problem because it requires to average over a number of disorder realisations to extract the  properties of the system.
This motivated us to develop an efficient numerical approach to simulate the dynamics of generic open many-body quantum systems in presence of disorder. 

For clean homogenous systems, a few theoretical and computational methods have been put forward in the last years to simulate the physics of open systems under generic nonequilibrium conditions, such as renormalisation group calculations exploiting the Keldysh formalism \cite{Sieberer2016,Maghrebi16}, corner-space renormalization \cite{FinazziPRL15}, permutation-invariant solvers \cite{ShammahPRA2018}, full configuration-interaction Monte Carlo \cite{NagyPRA18}, tensor-network techniques \cite{MascarenhasPRA2015,CuiPRL2015,Jaschke2019,WernerPRL2016,OrusNat17} and cluster methods \cite{BiellaNLCE2018,Jin16}. 

A convenient way to model open quantum systems is via stochastic quantum trajectories \cite{carmichael1998statistical,PlenioRMP1998,GirvinPRA2008,Daley2014}, which can be used to compute expectation values and to reveal important dynamical features.
According to the specific stochastic protocol, pure quantum states evolve according to stochastic jump-diffusion processes. The density matrix of the system can be retrieved by a proper average over an ensemble of trajectories.
Quantum trajectories can also be combined with many-body methods, such as those based on matrix product states \cite{AngelakisNJP2012}, renormalization group approaches \cite{FinazziPRL15}, Gutzwiller mean-field \cite{Wim2018}, semiclassical approaches \cite{MorigiPRA2017}, or with an analytical ansatz on the nature of the states \cite{Verstraelen2018}.
Critical phenomena in extended bosonic \cite{Vicentini2018} and spin \cite{Masca2017,Rota2018} lattices have been studied thanks to the Monte Carlo sampling of those trajectories. The main computational advantage of this approach is that one only needs to evolve pure states instead of the full density matrix of the system, reducing the memory cost.

In this paper we propose an optimal stochastic trajectory approach to efficiently deal with disordered systems.
We show how the additional computational cost needed to evaluate disorder configurations can be strongly reduced by sampling the disorder distribution with stochastic trajectories, evolving according to different disorder configurations.
We prove, under general assumptions, that our approach provides the best estimate of the exact disorder-averaged expectation values at a fixed computational cost.

Our method can be applied to generic open quantum many-body systems. In the present work, we focus on the paradigmatic driven-dissipative Bose-Hubbard model \cite{Vicentini2018,Foss-FeigPRA17,LeBoitePRL13} on a square lattice. 
We test our method against exact simulations and we combine it with several trajectory protocols to explore both the strongly and the weakly interacting regime.

In the strongly interacting limit, we studied the impact of disorder on the spectral properties of a driven-dissipative fermionized gas of bosons in one dimension  \cite{GirardeauJMP60}.
Tuning the frequency of the pump, one can excite different many-body resonances which, in the clean system, respect a selection rule based on the conservation of the quasi-momentum \cite{CarusottoPRL09}.
We investigate how the addition of disorder implies the violation of this rule and allows to access new resonances with spectroscopic experiments.

The driven-dissipative Bose-Hubbard model in two dimensions is known to also undergo a first-order dissipative phase transition from a low- to a high-density phase as the driving strength is increased \cite{Vicentini2018,Foss-FeigPRA17}. 
At the critical point, the system is bistable and the whole lattice site population jumps simultaneously between two metastable states \cite{Minganti2018}.
We study the fate of this transition in the presence of on-site disorder in the weakly interacting regime.
For closed systems, the effect of disorder on a classical \cite{Proctor2014,Imry1975,Imry1979} and quantum \cite{Hrahsheh2012} first-order phase transition is not as well known as the second-order.
In the case of a dissipative criticality this issue, to the best of our knowledge, has never been addressed so far. 

Beyond the fundamental theoretical interest, the method and the findings of our work are relevant for ongoing experiments in extended driven-dissipative photonic lattices \cite{HartmannJOP16,NohReview17}. 
Despite the impressive degree of controllability at single and few-site level, the presence of inhomogeneities when assembling several fundamental units is, at present, unavoidable. 
In the case of photonic lattices realised via semiconductor cavities and superconducting circuits the main source of disorder is local, i.e. the bare cavity frequency varies from site to site.  
So that, in order to predict the outcome of experiments (often analyzed by means of quantum and homodyne trajectories) one need to access the role of inhomogeneities and its relation to collective phenomena.

The paper is organized as follows:
In Sec. \ref{sec:disorder} we introduce the general framework to study disordered open quantum systems while in Sec. \ref{sec:method} we describe our method. The driven-dissipative Bose-Hubbard model is introduced in Sec. \ref{sec:bh-model}, together with numerical verification of the optimality of our strategy. We then exploit the optimal stochastic unraveling method to investigate the strongly interacting limit of this model in Sec. \ref{sec:strongly-interacting}, and the first-order dissipative transition in the weakly interacting regime in Sec. \ref{sec:weakly-interacting}. Finally, in Sec. \ref{sec:conclu} we draw our conclusions and discuss future perspectives.

\section{Disordered open quantum systems}
\label{sec:disorder}

In this paper we will consider quantum systems coupled to memoryless environments which leads to a Markovian dynamics for the system density matrix $\hat\rho$.
In this case, the dynamics is governed by a master equation in the Lindblad form ($\hbar=1$) \cite{BreuerBook07}
\begin{equation}
	\label{eq:lindblad-clean}
	\partial_t\orho(t) = \mathcal{L}\orho = -i \comm{\hat{H}}{\orho} + \sum_j\frac{\gamma_j}{2}\mathcal{D}[\hat{L_j}]\orho(t),
\end{equation}
where $\mathcal{L}$ is the Liouvillian superoperator. The commutator on the right hand side of Eq.\eqref{eq:lindblad-clean} accounts for the unitary part of the dynamics ruled by the Hamiltonian $\hat{H}$ while the incoherent dissipative processes are given by 
\begin{equation}
\mathcal{D}[\hat{L_j}]\orho(t) = 2\hat{L}_j\orho\hat{L}^\dagger_j-\acomm{\hat{L}^\dagger_j \hat{L}_j}{\orho},
\end{equation}
where $\hat{L}_j$ is the set of the jump operators representing different dissipative channels acting at a rate $\gamma_j$. 
At long times, the system will approach the steady-state $\hat\rho^{\rm SS}=\lim_{t\to\infty}\rho(t)$ (and thus $\partial_t\hat\rho^{\rm SS}=0$) regardless of the initial conditions. In what follows we will consider situations where $\hat\rho^{\rm SS}$ is guaranteed to be unique \cite{Nigro2018,AlbertPRA14}.

In a system with static disorder, we take into account random variations of the parameters describing the system, which we encode in the vector of random variables $\vb{w}$. 
Those inhomogeneities satisfy a certain probability distribution $p_D(\vb{w})$, which will determine how likely a certain configuration $\vb{w}$ is to occur. 
The open system dynamics of the disordered system is obtained by adding to the Lindbladian of the clean system $\mL_{\rm clean}$ a second Lindblad term $\mL_{D}(\vb{w})$, which accounts for both coherent and incoherent disorder-induced processes

\begin{equation}
\label{totliouv}
	\mL(\vb{w}) = \mL_{\rm clean} + \mL_{D}(\vb{w}).
\end{equation}

Let us note that $\mL(\vb{w})$ is still a well-defined Lindbladian according to Eq.\eqref{eq:lindblad-clean} and can be formally integrated to give the time-evolution of the density matrix for the system in the presence of static disorder,
\begin{equation}
	\label{eq:formal-rho-time-evo}
	\orho(t, \vb{w}) = {\rm e}^{\mL(\vb{w}) t} \orho_0,
\end{equation}
where $\orho_0$ is the state at $t=0$. The expectation value of an observable $\hat{O}$ at a given time $t$ for a specific disorder realization $\vb{w}$ reads as
\begin{equation}
	\langle\hat{O}\rangle(t,\vb{w}) = \Tr\left[ \hat{O}\rho(t, \vb{w}) \right].
\end{equation} 
In order to access the statistical properties of the system we need to take the \textit{average} of observables over the possible disordered configurations $\vb{w}$, weighted by their probability $p_D(\vb{w})$. We will refer to this average as the \textit{configuration} average, and will represent it with an overline $\overline{\bullet}$.
 Dropping the time-dependance we get:
\begin{align}
\label{eq:disorder-average-def-fi}
	\overline{\langle\hat{O}\rangle} &= \int d\vb{w} \ p_D(\vb{w}) \ \langle\hat{O}\rangle(\vb{w}) \\ 
			  &= \lim_{R\rightarrow\infty}\frac{1}{R}\sum_{i=1}^R \langle\hat{O}\rangle(\vb{w}_i),
			  \label{eq:disorder-average-def}
\end{align}
where, in the second line, we introduced the sum over the set of $R$ disorder configurations $\{\vb{w}_i\}_{i=1,\dots,R}$ sampled according to the distribution $p_D(\vb{w})$. 

Note that it is always possible to define a disorder-averaged density matrix as
\begin{equation}
\label{rhoave}
\overline{\orho}(t) = \left[\int d\vb{w} \ p_D(\vb{w})e^{\mL(\vb{w}) t }\right]\orho_0,
\end{equation}
such that $\Tr[\overline{\orho}(t) \hat{O} ] = \overline{\langle\hat{O}\rangle}$ for any given observable $\hat{O}$.
Through Eq.(\ref{rhoave}), it is also possible to define the averaged (or quenched) propagator  
\begin{equation}
\label{lave}
\exp[\mL_{\rm ave}t]=\int d\vb{w} \ p_D(\vb{w})e^{\mL(\vb{w}) t },
\end{equation}
which governs the averaged dynamics of the system.
This would allow us to obtain directly the averaged dynamics of the system in the particular cases when the formal 
expression \eqref{lave} can be exactly integrated.

\section{Optimal stochastic unraveling of disordered open quantum systems}
\label{sec:method}

\begin{figure}
	\includegraphics[width=\columnwidth,trim={0 1.3cm 0 0}]{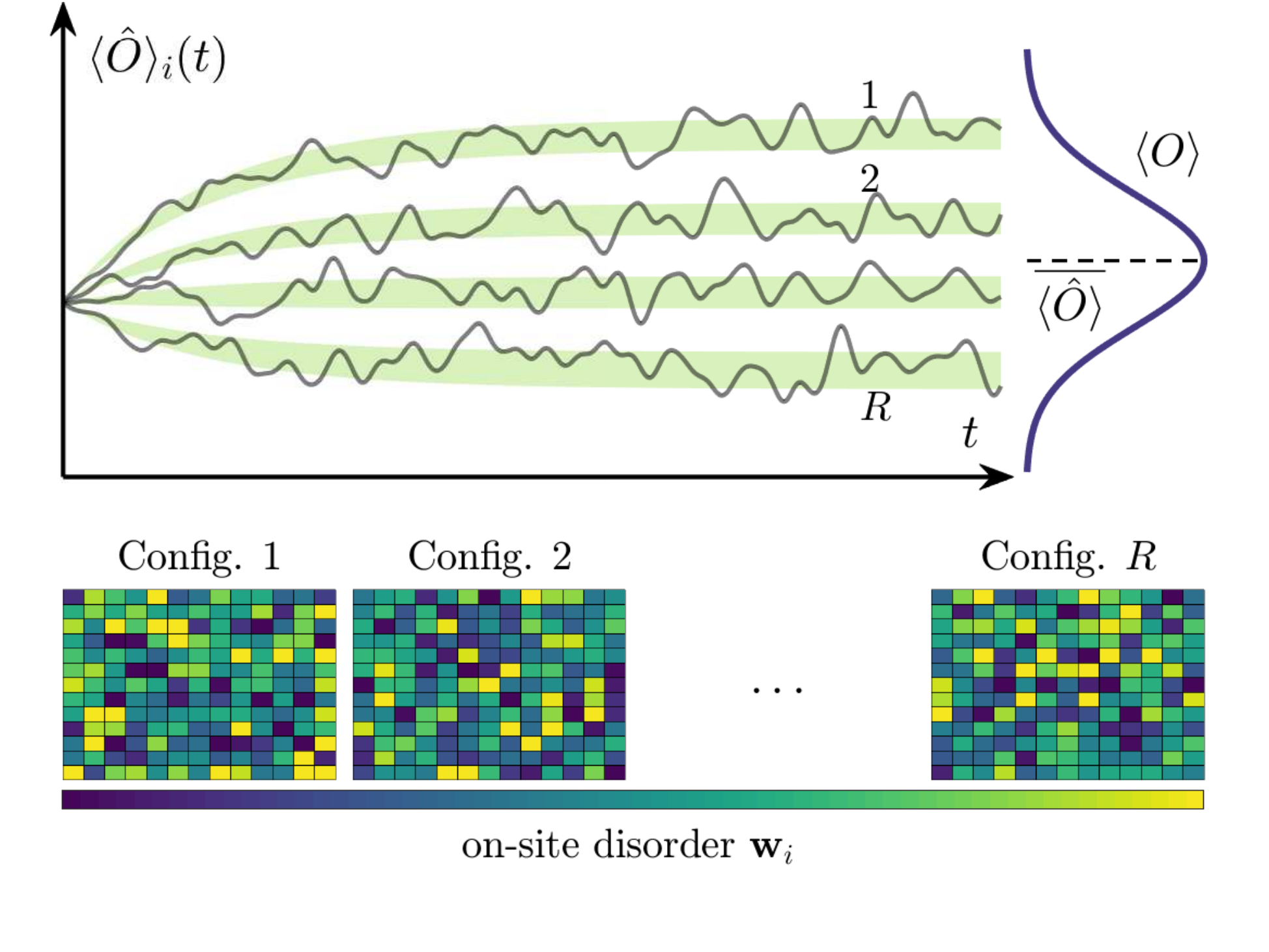} 
	\caption{Sketch of the optimal sampling method for a disordered open system, whose dynamics is unraveled in terms of individual stochastic trajectories.
	The configuration-averaged expectation value $\overline{\langle\hat{O}\rangle}(t)$ of an observable is obtained by averaging its value $\braket{\hat{O}}_i(t)$ over an ensemble of trajectories $i = 1,2, \dots R$, where each individual trajectory is evolved in the presence of a different disorder configuration $\vb{w}_i$.
	}
	\label{fig:sketch}
\end{figure}

In this section we present our approach, sketched in Fig.\ref{fig:sketch}, to the problem of disordered open quantum systems. 
In Sec. \ref{ssec:method-clean} we briefly review trajectory-based methods for clean systems.
In Sec. \ref{ssec:single-trajectory-method} we introduce our extension to efficiently compute disorder-averaged expectation values.
A numerical demonstration of the optimality of our method will be given in Sec. \ref{sec:bh-model}.

\subsection{Stochastic trajectories} 
\label{ssec:method-clean} 

Let us start by briefly reviewing the main ideas of the stochastic trajectory approach to clean open quantum systems ($\mathcal{L}_{D}=0$).
To compute numerically the state of the system at a given time, one needs to evolve the density matrix of the system according to Eq. \eqref{eq:lindblad-clean}.
This method is limited by very high memory requirements and it becomes unfeasible as the number of sites is large enough. 
Indeed, one need to store and evolve the full density matrix $\hat{\rho}$ which is a Hermitian matrix of dimension  $d^{N}\times d^{N}$ for a system made of $N$ sites with local Hilbert space having dimension $d$.

As already mentioned in the introduction, a possible approach with a lower memory cost involves the stochastic unravelling of the density matrix by means of trajectories evolving according to a given protocol \cite{Daley2014}. 
The unravelling can be formally written as an integral over the space of trajectories 
\begin{equation}
\label{rhounrav}
\hat\rho = \int_\mathcal{H} d\psi \ p(\psi)\ket{\psi}\bra{\psi},
\end{equation}
where we omitted the time-dependance. 
Such procedure allows us to interpret the density matrix at each time $t$ as the mixture of several pure states $\ket{\psi}$, and to recast the master equation \eqref{eq:lindblad-clean} into a jump-diffusion stochastic differential equation evolving a set of pure states \cite{carmichael1998statistical}. 
In practice, the distribution $p(\psi)$ is sampled by independently evolving a finite number of trajectories. 
As this is a stochastic sampling of the distribution, the error of the estimated expectation value $\hat{O}$ scales as \footnote{This assumes that all trajectories are independent. Because the initial state is fixed, for small times below the correlation time the trajectories are not independent variables, and a covariance correction to the variance should be included through the substitution $\Var_\psi\rightarrow\Var_\psi + \Cov_\psi$. As this is related to the correlation time, it must decay exponentially in time. Note that this only affects the efficiency of the sampling, but not the expectation value, which is unaffected.}
\begin{equation}
\label{eq:error-clean}
	\epsilon_{\rm clean} = \alpha\ \sqrt{\frac{\Var_\psi[\hat{O}]}{T}},
\end{equation}
where $T$ is the number of trajectories and $\Var_\psi[\hat{O}]$ is the statistical variance of the observable across the trajectories which depends on the specific evolution protocol and on the observable $\hat{O}$. The prefactor $\alpha=\epsilon/\sigma$ arises from the definition of the absolute deviation $\epsilon = \langle|\hat{O} - \langle\hat{O}\rangle|\rangle$, as opposed to that of the standard deviation $\sigma=\sqrt{\langle\hat{O}^2-\langle\hat{O}\rangle^2\rangle}$. As it is obvious from the definition, both are measures of the deviation (spread) of the distribution $\mathcal{O}$, but while the first has a more \textit{direct} interpretation when sampling, the latter has much better mathematical properties. It is possible to show that $\alpha\in[0,1]$, and we will simply treat it as a trivial proportionality factor with a value usually around that of the normal distribution, namely $\alpha=\sqrt{2/\pi}$ \cite[pgs.~81-85 and 113]{Kenney1947Book}.

\subsection{Stochastic trajectories for disordered systems}
\label{ssec:single-trajectory-method}

We can now introduce our method to perform an efficient estimate of disorder-averaged expectation values. 
Our aim is to efficiently compute the expectation value defined by Eq. \eqref{eq:disorder-average-def-fi} when the density matrix is unravelled stochastically. 
By inserting Eq.\eqref{rhounrav} into Eq.\eqref{eq:disorder-average-def-fi} we get:
\begin{equation}
	\overline{\langle\hat{O}\rangle} = \int d\vb{w} \ p_D(\vb{w})\int d\psi \ p(\psi|\vb{w}) o(\psi),
\end{equation}
where $p(\psi|\vb{w})=p(\psi,\vb{w})/p_D(\vb{w})$ is the conditional probability of obtaining the state $\ket{\psi}$ given the disorder configuration $\vb{w}$ and $o(\psi) = \bra{\psi(\vb{w})} \hat{O}\ket{\psi(\vb{w})}$.
Sampling the disorder-averaged expectation value can therefore be interpreted as performing the sampling of multivariate distribution $p(\psi,\vb{w})$ which depends on $\psi$ and $\vb{w}$. 

There exist several ways to perform such a sampling. 
Indeed, we have freedom to choose how many trajectories we want to evolve according to a given disorder configuration.
In the general case one evolves $T$ trajectories for each of the $R$ disorder realizations for a total computational cost $\mathcal{C}=R\times T$. 
We will show that, when computing disorder-averaged expectation values, the most computationally efficient sampling strategy is realized for $T=1$, i.e. one needs to evolve only one trajectory for each disorder configuration.
This strategy is sketched in Fig. \ref{fig:sketch}.
In this case, the error committed when computing the expectation value is
\begin{align}
	\epsilon^2 &=\alpha^2\ \frac{\overline{\langle \hat{O}^2\rangle} - \left(\ \overline{\langle \hat{O}\rangle}\ \right)^2}{R} \\
	&=\alpha^2 \ \frac{\overline{\langle \hat{O}^2\rangle - \langle{\hat{O}\rangle}^2} + \left[\overline{\langle\hat{O}\rangle^2} -\left(\ \overline{\langle \hat{O}\rangle}\ \right)^2\right]}{R}.
	\label{eq:terms-error}
\end{align}
The first term $\overline{\langle \hat{O}^2\rangle - \langle{\hat{O}\rangle}^2} = \overline{\Var_\psi[{\hat{O}}]}$ is the disorder-averaged statistical variance of the observable [see Eq.\eqref{eq:error-clean}] and accounts for the uncertainty due to the trajectory protocol. 
This is the same contribution found in the clean case described by Eq.\eqref{eq:error-clean}.
The second term $\overline{\langle\hat{O}\rangle^2} -\left(\ \overline{\langle \hat{O}\rangle}\ \right)^2 = \Var_D[\langle\hat{O}\rangle]$, instead, is the variance of the expectation value of the observable $\langle\hat{O}\rangle$ due to the presence of disorder. 
Exploiting these identifications, we can rewrite Eq.\eqref{eq:terms-error} in a more compact form as
\begin{equation}
\label{T1error}
	\epsilon=\alpha\ \sqrt{\frac{\overline{\Var_\psi(\hat{O})} + \Var_D(\langle\hat{O}\rangle)}{R}},
\end{equation}
which is valid when sampling one trajectory for each disorder realization. 
The generalization of Eq.\eqref{T1error} to the $T>1$ case is obtained by performing the substitution $\Var_\psi[\hat{O}]\rightarrow\Var_\psi[\hat{O}]/T$ which reflects the reduction in the statistical error when evolving multiple trajectories.
After some algebra, we obtain the general formula for the error as a function of $R$ and $T$
\begin{equation}
\label{error_general}
	\epsilon(R,T)=\alpha\ \sqrt{\frac{\overline{\Var_\psi(\hat{O})} + T\Var_D(\langle\hat{O}\rangle)}{RT}}.
\end{equation}
We point out that in the absence of disorder, $ \Var_D=0$ and the only term playing a role is the uncertainty associated with the trajectories. In this case, the distinction between different disorder configurations becomes meaningless, and   Eq.\eqref{error_general} reduces to Eq.~\eqref{eq:error-clean} where the total number of trajectories is $T\rightarrow\mathcal{C}$. Equivalently, when integrating the master equation exactly for $R$ different configurations of $\vb{w}$ one obtains a similar formula where the contribution of the trajectories is absent ($\Var_\psi=0$) and only $\Var_D$ contributes to the error.

From Eq. \eqref{error_general} it is easy to deduce that the most efficient sampling strategy, i.e. the choice of $T$ which minimizes the error for a fixed computational cost $\mathcal{C}=R\times T$, is realised for $T=1$. 
We remark that the only assumption we made to obtain Eq.\eqref{error_general} is to consider that the dynamics of the $T$ trajectories we evolve for each disorder realization is not correlated \footnote{If the evolution is started from a pure state, then the trajectories will be correlated for times $t<\tau$ below the correlation time. Nevertheless, our method still yields correct results, but the error will scale slightly worse, as there are cross-correlations that are harder to average out.}, as required for an effective stochastic unraveling \cite{gardiner2004quantum}. Thus Eq.\eqref{error_general} holds for arbitrary evolution protocols and disorder distributions.

\section{Numerical applications to the driven-dissipative Bose-Hubbard model}

\label{sec:bh-model}
\label{model}

As a first application of the optimal stochastic unraveling of disordered open quantum systems, we study the driven-dissipative Bose-Hubbard model with on-site static disorder. 
The coherent part of the evolution (in the frame rotating at the driving frequency and setting $\hbar=1$) is ruled by the following Hamiltonian
\begin{equation}
	\ham_{\rm BH} = \sum_{j=1}^L\left[-\Delta_j\oad_j\oa_j + \frac{U}{2}\hat{a}^{\dagger  2}_j\oa_j^2 + F\left(\oad_j+\oa_j\right)\right] - J\sum_{\langle i,j\rangle} \oad_i\oa_{j}, 
	\label{eq:hamiltonian-BH}
\end{equation}
where $\Delta_j=\omega_p-\omega_j$ is the local frequency detuning ($\omega_p$ and $\omega_j$ being the drive and the local bare frequency, respectively), $U$ is the on-site interaction, $F$ the strength of the coherent drive (the phase is set in such a way that $F\in\mathbb{R}$) and $J$ sets the hopping rate between nearest-neighbouring sites. 
The local detunings $\{\Delta_j\}$ are assumed to be random variables obeying a gaussian distribution $\mathcal{N}(\Delta_0,W)$ with mean $\Delta_0$ and variance $W^2$. 
Dissipative processes, i.e. local boson leakage, are modelled with a set of local jump operators $\hat{L}_j=\oa_j$ ($j$ being the site index) with uniform rate $\gamma$ leading to the following expression for the master equation 
\begin{equation}
	\partial_t\orho (t) = -i\comm{\ham_{\rm BH}}{\orho} + \frac{\gamma}{2}\sum_{j=1}^L\left[2\oa_j\orho\oad_j  - \acomm{\oad_j\oa_j}{\orho} \right].
	\label{eq:lindblad-me}
\end{equation}

\begin{figure}[t]
	\includegraphics[width=\columnwidth]{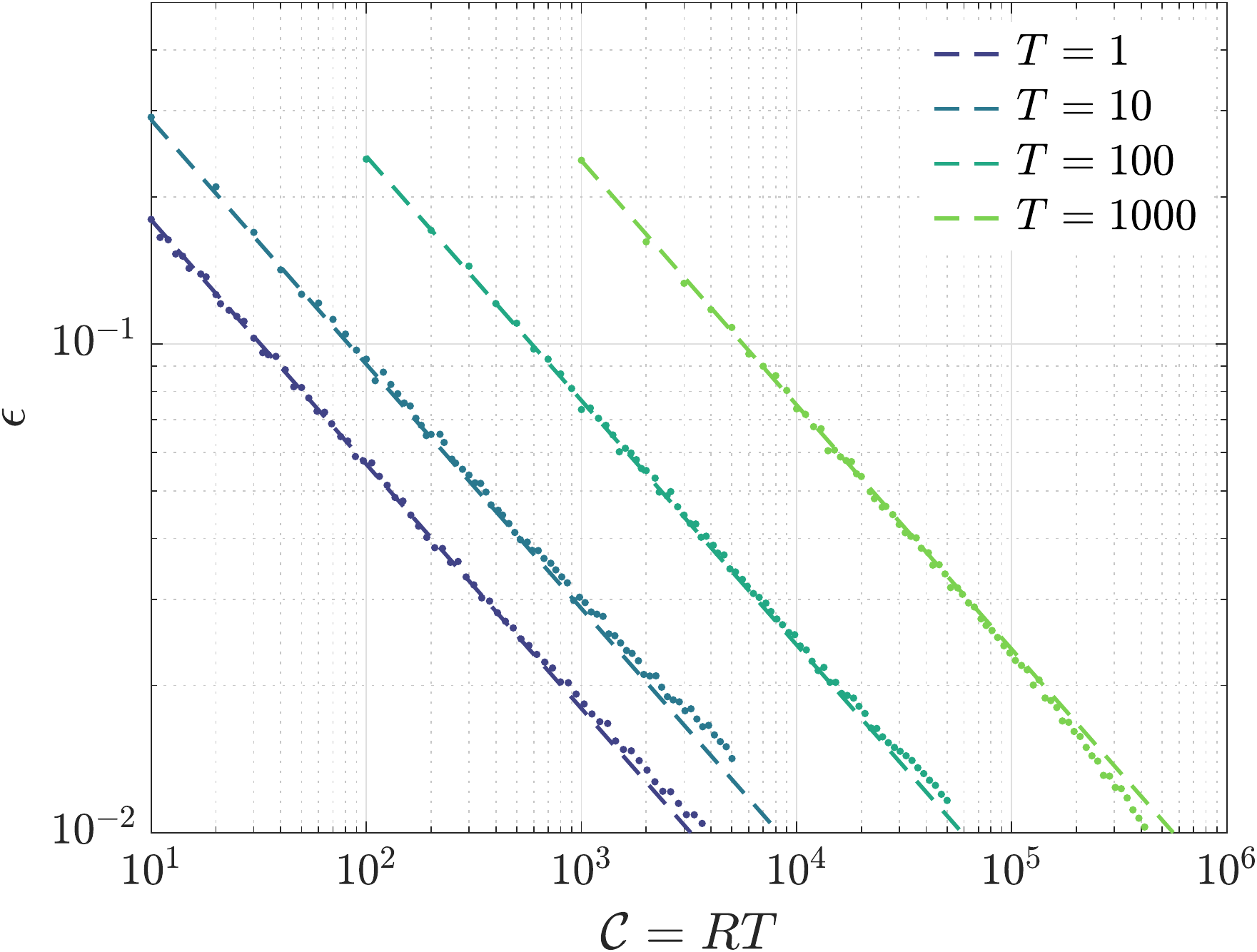} 
	\caption{Mean absolute error $\epsilon=|n_{ss}-n_{\rm exact}|/2$ as a function of the computational cost $\mathcal{C}$, for the average number of bosons in a driven-dissipative Bose-Hubbard chain made of $5$ sites with $U=1.0\gamma$, $F=2.0\gamma$, $J=0.5\gamma$ and $\Delta=\mathcal{N}(1.0, 0.5)$. The dashed lines are predicted by $\epsilon(\mathcal{C}/T, T)$ as defined in Eq.\eqref{error_general}, while dots are numerical data. The cases of $T=1, 10, 100, 1000$ trajectories are considered. Numerical data has been computed using a combination of \textit{QuTiP} \cite{qutip2012,qutip2013} and \textit{QuantumOptics.jl} \cite{quantumopticsjl}. }
	\label{fig:small1}
\end{figure}

Let us start by presenting a numerical verification of the optimality of our sampling method by considering a driven-dissipative Bose-Hubbard chain consisting of $L=5$ sites with periodic boundary conditions and interaction strength $U=\gamma$.  
First, we have computed the observables via a brute force integration of the master equation averaged over $2000$ disorder configurations, sufficient to reduce the statistical error below $5\cdot 10^{-3}$. We take the value of the average boson density $\hat{n}=\sum_{j=1}^L\oad_j\oa_j/L$ computed in this way as the \textit{exact} value $n_{exact}$ against which we will compare the sampling performed with trajectories.

We then unravel the dynamics of the master equation (Eq.\ref{eq:lindblad-me}) in terms of quantum trajectories \cite{Carmichael2008} within the so-called wavefunction Monte Carlo technique \cite{DalibardPRL1992, WFMCMolmer1993}. In this framework, a pure quantum state $\ket{\psi(t)}$ is evolved according to a non-Hermitian Hamiltonian in the presence of stochastic quantum jumps. This jump-differential equation reads

\begin{multline}
\frac{d\ket{\psi(t)}}{dt} = -i\left(\hat{H}(\vb{w}) - i\gamma\sum_j \hat{a}^\dagger_j \hat{a}_j\right)\ket{\psi(t)} + \\ + \sum_j N_j(t)\ \hat{a}_j \ket{\psi(t)},
\end{multline}
where $N_j(t)$ is a Poissonian counter which is equal to 1 with probability $p_j(t) = \gamma\expval{\hat{a}_j^\dagger \hat{a}_j}{\psi(t)}/\ip{\psi(t)}{\psi(t)}$ and 0 otherwise, encoding the stochastic nature of quantum jump trajectories.  
Note that the above equation does not preserve the norm of $\ket{\psi(t)}$, which should be re-normalized when computing observables.
The disorder configuration is encoded in the vector of on-site energies $\vb{w}=\{\Delta_i\}$. 
For every configuration $\vb{w}$ we evolve $T=1,\ 10,\ 100,\ 1000$ trajectories.
Considering $R$ different disorder configurations, we extract the steady-state average boson density $n_i$ for every $i=1,\dots,\mathcal{C}=RT$ trajectory. 

In Fig.\ref{fig:small1} we show the mean absolute error $\epsilon(R,T)=\frac{1}{\mathcal{C}}\sum_{i=1}^{\mathcal{C}} |n_i - n_{exact}|$ computed both numerically and analytically (Eq.\eqref{error_general} with $\alpha=\sqrt{2/\pi}$), finding a perfect agreement. 
The fact that at a fixed computational cost $\mathcal{C}$ the lowest sampling error $\epsilon$ is achieved by considering $T=1$ further confirms the optimality of our method.



In what follows we will adopt this method to compute disorder averages and study the effect of on-site disorder on a first-order dissipative phase transition in the driven-dissipative Bose-Hubbard lattice. 

\subsection{Strongly interacting (fermionized) bosons in 1D chains}
\label{sec:strongly-interacting}

In this section we will focus on the strongly interacting limit $U/J \gg 1$, where the boson-boson interaction is strong enough to forbid the double occupation of each site.
In this hard-core limit, the many-body Hamiltonian \eqref{eq:hamiltonian-BH} can be exactly diagonalized in one spatial dimension with periodic boundary conditions via {\it fermionization} of the bosonic wavefunction. 
Indeed, any $N$-body bosonic wavefunction $\Psi_B(i_1, i_2, \dots, i_N)$ ($i_1, i_2, \dots, i_N$ being the positions of the $N$ particles) can be one-to-one mapped onto a fermionic one \cite{GirardeauJMP60}.
For $N$ bosons, the wavefunction in real-space representation is given by 
\begin{equation}
\Psi_B(i_1, i_2, \dots, i_N)=\prod_{k<j}^{N} {\rm sgn} (i_k-i_j) \ \Psi_F(i_1, i_2, \dots i_N), 
\end{equation}
where $\prod_{k<j}^{N} {\rm sgn} (i_k-i_j)$ ensures that $\Psi_B$ respects the bosonic statistics.
In the limit of vanishing pumping (i.e., $F=0$), the eigenstates of \eqref{eq:hamiltonian-BH} can be simply labelled by the occupation number of single-particle eigenstates of a free-fermion system.
The notation $\ket{k_1,\dots, k_N}$ indicates the fermionic eigenstates obtained via the symmetrization of the wavefunction with one particle in each $k_1, \dots, k_N$ orbital \cite{CarusottoPRL09}.
The energy of this state with $N$ particles, according to Eq.~\eqref{eq:hamiltonian-BH} is 
\begin{equation}\label{Eq:resonances}
\mathcal{E}(k_1, \dots, k_N)= - N \Delta_0  -2 J \sum_{i=1}^{N} \cos(k_i).
\end{equation}

In a spectroscopic experiment, as the pump frequency is varied, one do expect an $N$-body resonance to appear whenever $\Delta_0 = -2 J \sum_{i=1}^{N} \cos(k_i)/N$.
In particular, when the system is homogeneously driven, only many-body states with total momentum $\mathcal{K}=\sum_i^N k_i=0$ can be accessed. 
We also remark that, because the pump term $F(\hat{a}+\hat{a}^\dagger)$ only couples the vacuum with 1-body states $\ket{k_n}$ at first order in $F$, resonances with $N\geq2$ will only arise when the pump strength $F$ is non-perturbative with respect to $\gamma$. 

\begin{figure}
	\centering
	\includegraphics[width=\linewidth]{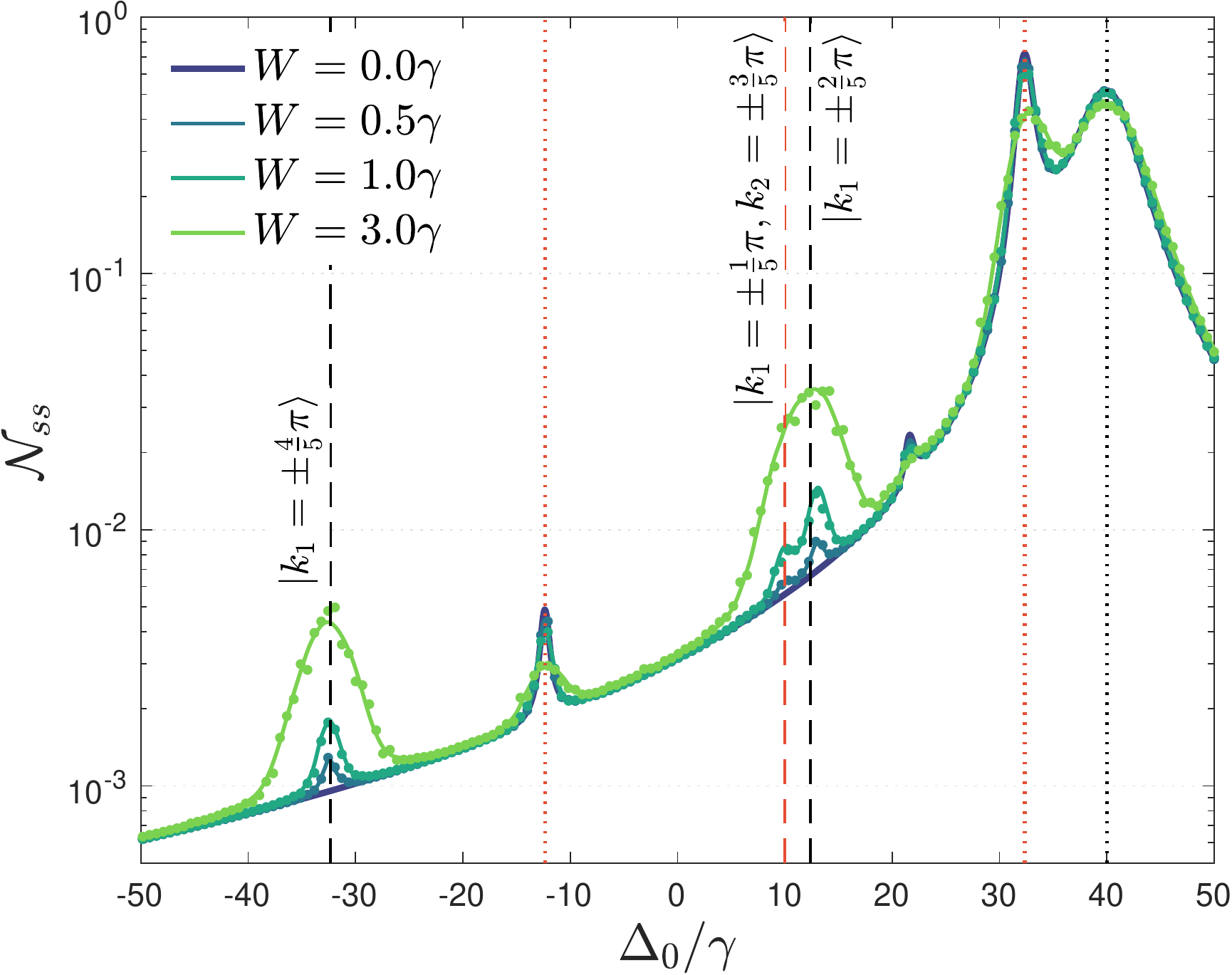}
	\\
	\includegraphics[width=\linewidth]{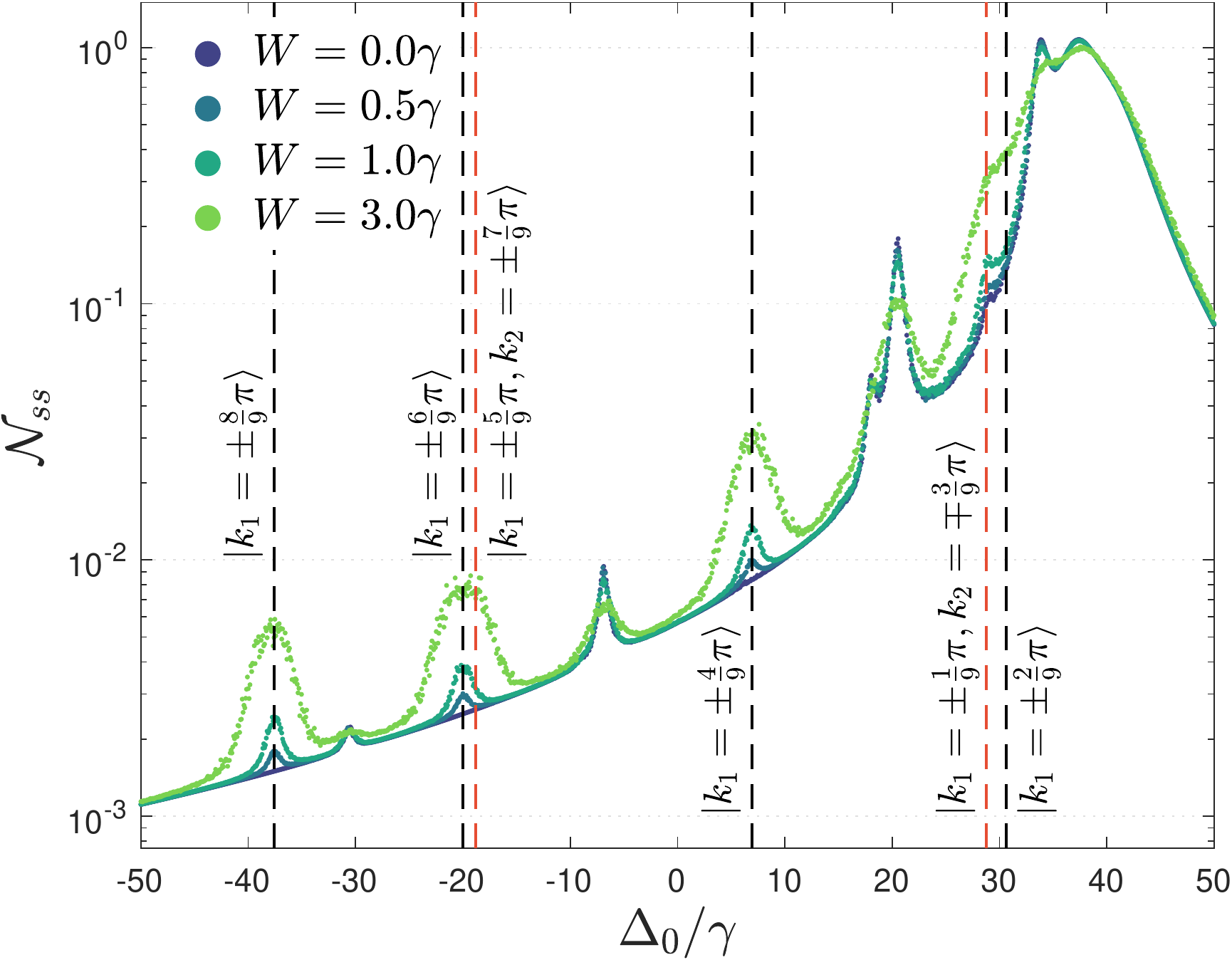}
	\caption{Mean boson occupation number as a function of the detuning $\Delta$ for a closed chain of 5 (top) and 9 (bottom) sites populated with hardcore bosons. Vertical dotted lines are resonances with $\mathcal{K}=0$, while dashed lines are resonances with $\mathcal{K}\neq0$. Black lines are $1-$particle resonances while orange lines (light gray) are $2-$particle.
	Top: Smooth curves have been obtained with brute force integration, while dotted points have been computed with the method presented in Sec.\ref{sec:method}. 
	Parameters: $F/\gamma=1.0$, $J/\gamma=20$.
	}
	\label{fig:resonances_5_cav}
\end{figure}

While the spectroscopy of the {\it clean} system has been studied in Refs.\cite{Biella2015,CarusottoPRL09}, here we want to characterize the effect of local disorder on the resonances of the fermionized system. 
To this aim, we studied the average population in the steady-state 
$\mathcal{N}_{ss}=\sum_j \braket{\hat{a}_j^\dagger \hat{a}_j}$, a witness of the presence of those $N-$particle resonances, as a function of the detuning $\Delta_0$. 
In Fig.~\ref{fig:resonances_5_cav} (a) and (b) we show $\mathcal{N}_{ss}$ for a chain of five and nine resonators, respectively. 
The position of one- and two-body resonances which respect the selection rule $\mathcal{K}=0$ is marked by black and orange (light gray) dotted lines, respectively.  
The position of the others one- and two-body resonances with $\mathcal{K}\neq0$ is denoted by dashed lines. 
In the panel (a) the continuous line has been obtained by averaging the exact solution obtained via exact diagonalization of the Liouvillian over $R=300$ disorder realizations for each value of $\Delta_0$.
To show the advantage of our method introduced in Sec.\ref{sec:method}, we computed the same observables through a wavefunction Montecarlo algorithm \cite{WFMCMolmer1993,CarmichaelPRA93,Daley2014} where every quantum trajectory evolves according a different realization of disorder. 
The data obtained with our method (filled dots) are in very good agreement with the exact simualtions, and can be obtained (for a grid of equal spacing) with a computational cost that is roughly $2^5=32$ times lower.

Let us now focus on the position of the resonances.
If $W=0$, only the resonances for which $\mathcal{K}=0$ are excited by the homogeneous drive. 
For small disorder (i.e. $W/\gamma<1$) the intensity of the $\mathcal{K}=0$ resonances is only affected perturbatively. 
However, one-body resonances with $\mathcal{K}\neq0$ start to be visible in the spectrum. 
Indeed, the disorder perturbatively changes the shape of one-body states allowing to populate them even with an homogeneous pump profile.
 
Among the several two-body $\mathcal{K}\neq0$ resonances (most of which are not shown), the one at $\Delta/\gamma=10\gamma$ is particularly interesting because it is the only one significantly populated. 
An intuitive explanation is obtained by considering that the drive couples one- and two-body states with the same total momentum $\mathcal{K}$. Because the coupling is inversely proportional to their energetic mismatch, only  $\ket{\mp\frac{1}{5}\pi,\ \pm\frac{3}{5}\pi}$  which has $\mathcal{K}=\pm\frac{2}{5}\pi$ couple to the nearby $\ket{\pm\frac{2}{5}}$ one-body state since the pump conserves the total momentum $\mathcal{K}$. 
For large disorder ($W/\gamma>1$) this one- and two-body resonances start to overlap, eventually merging into a unique broad peak for larger values of $W$.

\begin{figure*}
	\includegraphics[width=0.9999\textwidth]{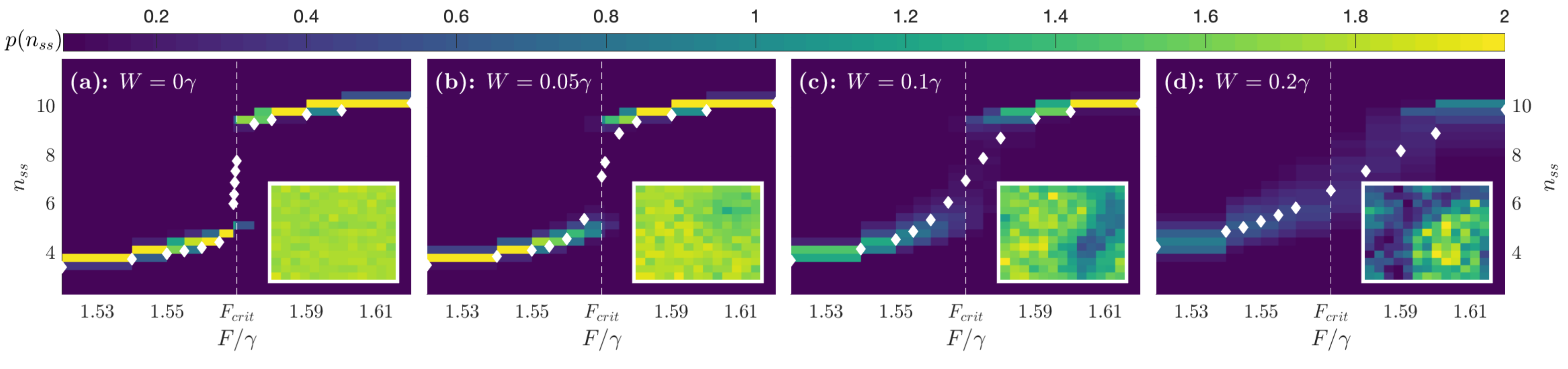}
	\caption{Probability distribution of the disorder-averaged steady-state population $n_{ss}$ for different amplitudes $W$ of the disorder. The white diamonds represent the steady-state expectation value $\expval{n_{ss}}$. 
The insets show a typical distribution of the site-dependent boson density at long times for a given disorder configuration.
Parameters are $J=0.225\gamma$, $\Delta=0.1\gamma$ in a $14\times 14$ lattice. 
Up to $1000$ trajectories, evolved for $t>4\tau$ where considered for each point.
For more details on the choice of those parameters refer to \cite{Vicentini2018}.  }
	\label{fig:pn-disorder}
\end{figure*}

This disorder-enhanced emergence of resonances is expected to become more relevant for longer chains, where more $\mathcal{K}\neq0$ resonances are now accessible. We exploited our technique to investigate a 9 site chain, which would otherwise be computationally much more demanding. 
The results are presented in the bottom panel of Fig.\ref{fig:resonances_5_cav}. 
In this case, all the five one-body resonances becomes visible (black vertical lines) and the nearby 
two-body states with $\mathcal{K}\neq0$  with the same $\mathcal{K}$ emerge.

We can conclude that, for the quite wide range of disordered considered, the resonances of the {\it clean} system are still observables when a spectroscopy of the system is performed.
Furthermore, some of the many-body resonances unaccessible because of the symmetry of the pump, emerge in presence of disorder.

\subsection{Role of disorder on a first-order dissipative phase transition in 2D lattices}
\label{sec:weakly-interacting}

In this subsection we will study the fate of a first-order criticality in a weakly interacting driven-dissipative Bose-Hubbard lattice with disorder.

The master equation Eq.\eqref{eq:lindblad-me} can be recasted into a third-order partial differential equation for the Wigner quasiprobability distribution \cite{carmichael1998statistical}.
Considering the regime of weak interactions, we can perform the truncated Wigner approximation by neglecting the third order terms, obtaining a well-defined Fokker-Planck equation. The time-evolution of the Wigner distribution can be efficiently sampled by solving the associated stochastic differential equation for the Wigner trajectories \cite{VogelPRA89}

\begin{multline}
	\frac{d\alpha_j}{dt} = \left[ -i (\Delta_j -  U (\vert \alpha_j \vert ^2 - 1) - \gamma/2 )\right] \alpha_j - \\ -i J \sum_{\langle j',j\rangle}  \alpha_{j'}  + i F + \sqrt{\gamma/2} \,\chi (t),
	\label{eq:wigner-trajectories}
\end{multline}
where $\expval{\chi(t)\chi^\star(t')} = \delta(t-t')$ and $\expval{\chi(t)\chi(t')} = 0$ is a delta-correlated noise.
Expectation values are obtained by averaging over a large number of such trajectories.
This approach has been shown to be very effective in the {\it clean} case for a   range of nonlinearities up to $U\lesssim0.5\gamma$ \cite{Vicentini2018}, allowing for the simulation of large lattices up to $22\times22$ sites. 
We solved Eq.\eqref{eq:wigner-trajectories} with a combination of recently developed stiff solvers \cite{Rackauckas17,Rackauckas18} implemented in \textit{Julia} \cite{Julia2017} within the \textit{DifferentialEquations.jl} package \cite{DifferentialEquationsjl2017}. 


In the homogeneous case ($W=0$), for some values of $\Delta_0$, the system is known to undergo a first-order phase transition in the steady-state from a low- ($F<F_{crit}$) to a high-density ($F>F_{crit}$) phase as the driving strength is increased \cite{Foss-FeigPRA17,Vicentini2018}. 
At criticality the density matrix is bimodal: the steady-state density matrix is the symmetric mixture of the two phases \cite{Minganti2018}. 
As a consequence, when $F\approx F_{crit}$, the system displays bistability: the whole lattice simultaneously jump between two metastable states (the low- and high-density phases). 
Such mechanism is the fingerprint of a dissipative first-order transition.
The transition rate depends on the asymptotic decay rate towards the steady state $\tau^{-1}$ \cite{VogelPRA88}, which vanishes in the thermodynamic limit leading to the {\it critical} slowing down in the system dynamics at the critical point \cite{Vicentini2018}. 
With this in mind, we exploited the stochastic unraveling method presented in this article to efficiently simulate disordered lattices and investigate the interplay of disorder and criticality.

\subsubsection{Boson density}
\label{ssec:density}

\begin{figure}
	\includegraphics[width=0.97\columnwidth]{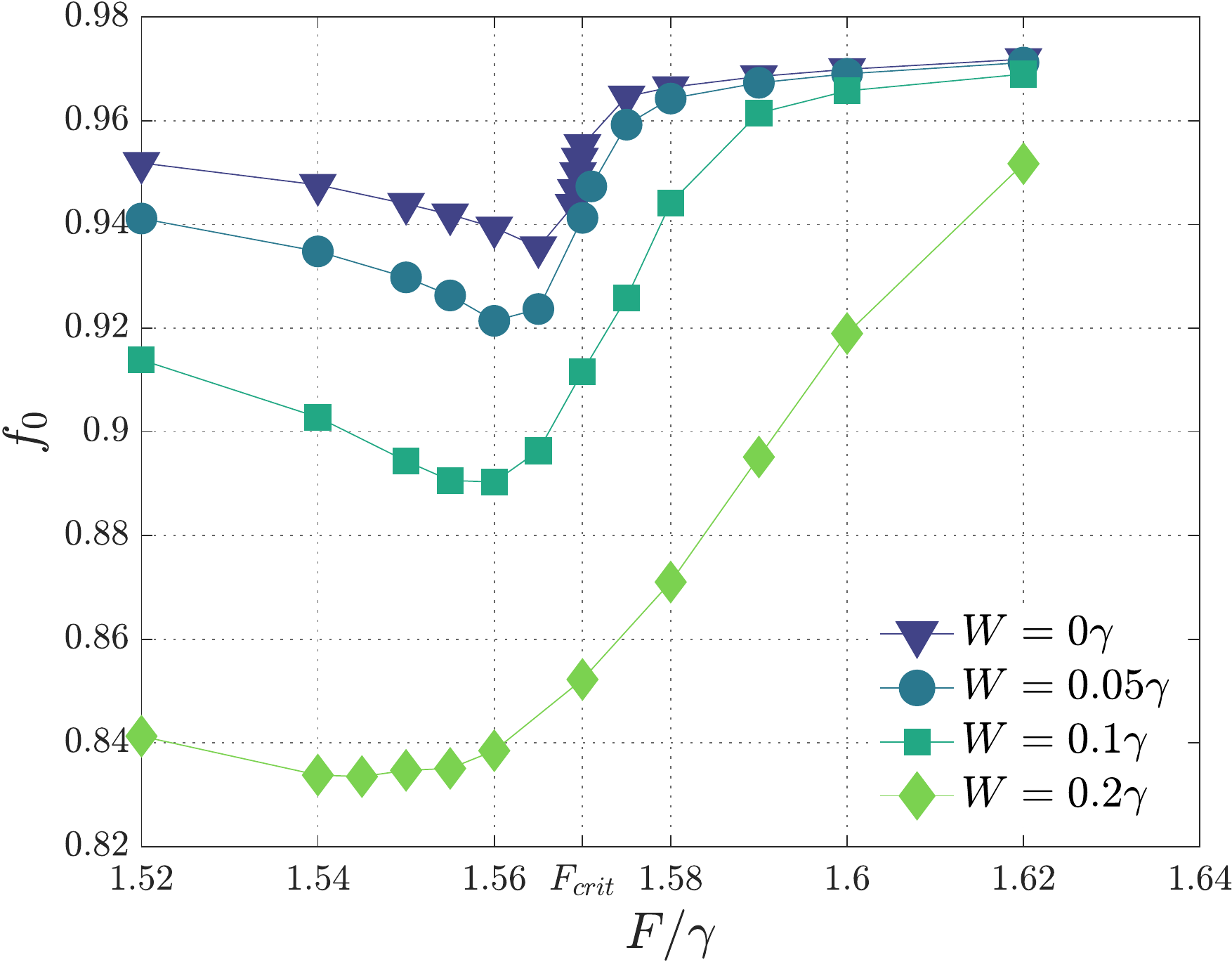}
	\caption{Fraction of bosonic population in the $\textbf{k}=0$ mode ($f_0=n_{\vb{k}=0}/n$) as a function of $F/\gamma$ for different disorder amplitudes in a $14\times 14$ lattice.
	Parameters as in Fig. \ref{fig:pn-disorder}.}
	\label{fig:nk0-disorder}
\end{figure}

Let us start our analysis by studying the bosonic population across the transition.
In Fig. \ref{fig:pn-disorder} we computed the probability distribution of the average steady-state density $n_{ss}$ for several values of the disorder $W$ on a $N=L\times L$ lattice with $L=14$.
In the insets we show typical snapshots of the site-resolved density along a single trajectory, where the formation of domains can be seen for stronger disorders.  
We note that the abrupt jump present in the clean case [see Fig. \ref{fig:pn-disorder}, panel (a)] becomes progressively smoother when influenced by the local disorder.
Interestingly, even for disorder distributions well within the single-site linewidth ($W\ll\gamma$) the transition is rounded and replaced by a smooth crossover.

The underlaying mechanism responsible for the suppression of the criticality is the depletion of the homogeneous $\vb{k}=0$ mode of the lattice which is macroscopically occupied across the transition for $W=0$.
In Fig. \ref{fig:pn-disorder}, we show the homogeneous fraction $f_0=n_{\vb{k}=0}/n=\sum_{i,j=1}^N\braket{\oad_i\oa_j}/\sum_{i=1}^N\braket{\oad_i\oa_i}$ in the steady state for different disorder strengths $W$.
Local inhomogeneities in the energy landscape force the system to populate $\textbf{k}\neq0$ modes, competing against the transition. 
We point out that the strongest depletion takes place for $F<F_{crit}$. 
Indeed, in the \textit{linear} regime the term that dominates the local energy is proportional to the random local detuning.
When $F>F_{crit}$ the leading term due to the nonlinearity $U$ grows as $n_{ss}^2$, making this phase less sensitive to the presence of disorder.

\begin{figure}
\includegraphics[width=0.96\columnwidth]{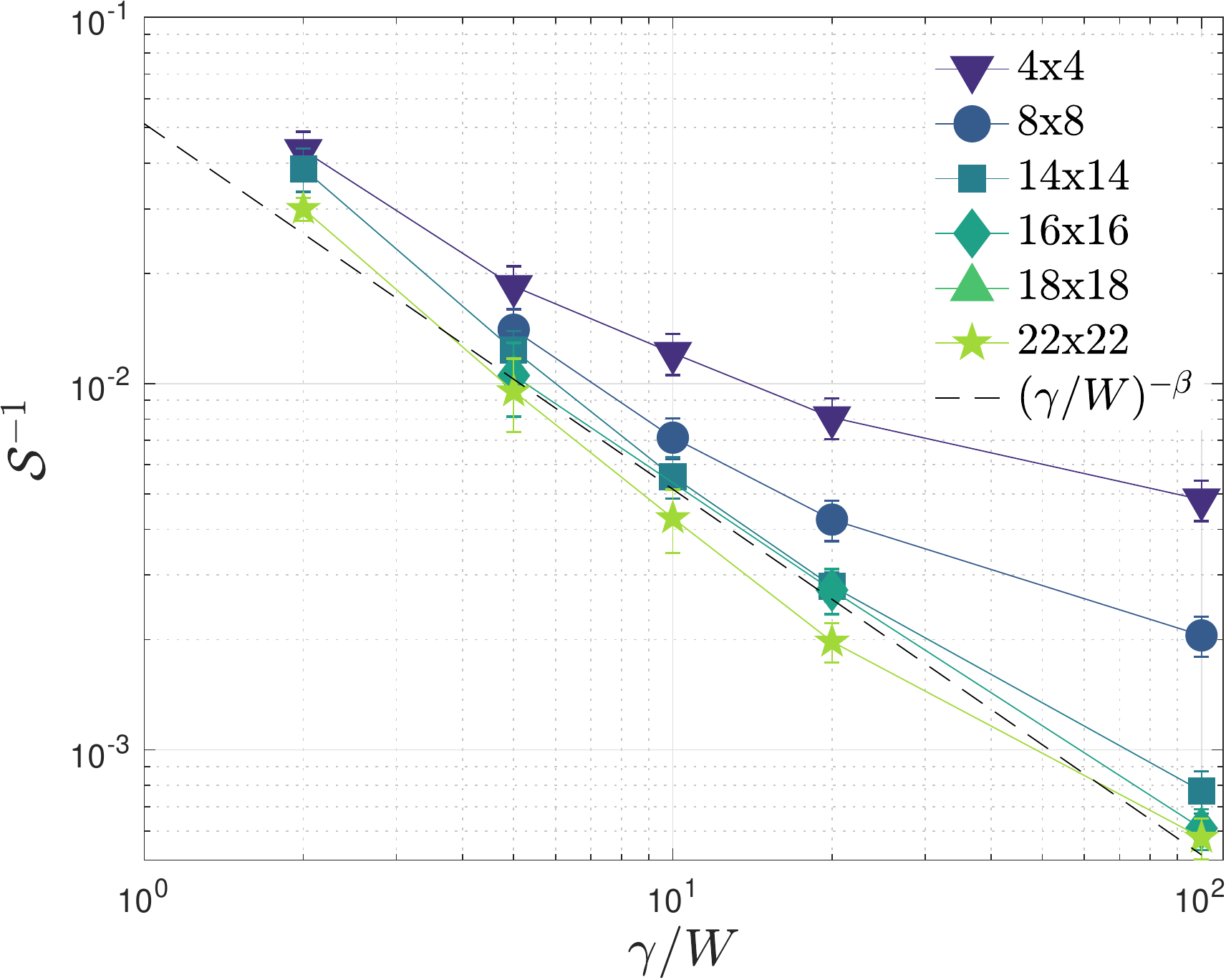}
\caption{Scaling of the inverse slope $\mathcal{S}^{-1}$ as a function of $\gamma/W$, for different lattice sizes. The dashed line is the power-law $\mathcal{S}^{-1} = (0.063\pm0.010)W^{0.98\pm0.05}$ obtained by fitting the $22\cross22$ data. 
Parameters as in Fig. \ref{fig:pn-disorder}.}	
\label{fig:scaling_S_W}
\end{figure}

To better characterise the effect of disorder on the criticality, we studied the behavior of the slope at the transition 
\begin{equation}
\left.\mathcal{S}=\frac{\partial n_{ss}(F/\gamma)}{\partial (F/\gamma)}\right|_{F=F^*},
\end{equation}
where $F^*$ is the value of $F$ for which $\mathcal{S}$ is largest.
In Fig. \ref{fig:scaling_S_W} we show $\mathcal{S}^{-1}$ as a function of $W^{-1}$ for different system sizes. 
As the size of the array is increased, the data progressively display a power-law behavior 
\begin{equation}
\label{eq:powlaw}
\mathcal{S}\sim W^{-\beta},
\end{equation} 
with $\beta=0.98\pm0.05$. 
This result proves that the discontinuous population jump occurring in the thermodynamic limit is smoothened by disorder. Thus only a perfectly homogeneous system ($W=0$) would display a true criticality $\mathcal{S}\to\infty$ when $L\to\infty$ if one drives the $\vb{k}=0$ mode. 
This does not exclude that the criticality could be restored by engineering a space-dependent driving field which couples more effectively to the modes of the disordered system.
 
The value of the exponent $\beta$ has been extrapolated by fitting the data for $L=22$.
Interestingly, any finite-size system made of $L\times L$ sites, follows the clean-system behavior in Eq. \eqref{eq:powlaw} up to a certain value $1/W^*$ which increases as $L$ is increased.
This is due to finite-size effects and, in particular, the relation between $W^*$ and $L$ is connected with the typical correlation length of the systems. 
To get further insight about this point, in the following section we study the behavior of the correlation functions at the transition in clean and disordered systems.

\subsubsection{Correlation functions}
\label{ssec:corr}

The rounding of the transition due to disorder is also witnessed by the connected part of the one-body correlation functions
\begin{equation}
\label{eq:corrfun}
g^{(1)}(r=|\vb{l}-\vb{m}|) = \braket{\oad_{\vb{l}}\oa_{\vb{m}}} - \braket{\oad_{\vb{l}}}\braket{\oa_{\vb{m}}},
\end{equation}
where $\vb{l}$ and $\vb{m}$ are the $2D$ coordinates of lattice sites.
In a coherently-driven system it is necessary to consider the connected part of the correlators in order not to account for the coherence imprinted by the external driving field \footnote{Given that $\braket{\oa_{\vb{i}}}\neq0, \ \forall \vb{i}$ because of the local driving term one would simply get $\lim_{|\vb{l}-\vb{m}|\to\infty}\braket{\oad_{\vb{l}}\oa_{\vb{m}}}\sim\braket{\oad_{\vb{l}}}\braket{\oa_{\vb{m}}}\to{\rm const}$.}.

\begin{figure}
\includegraphics[width=\columnwidth]{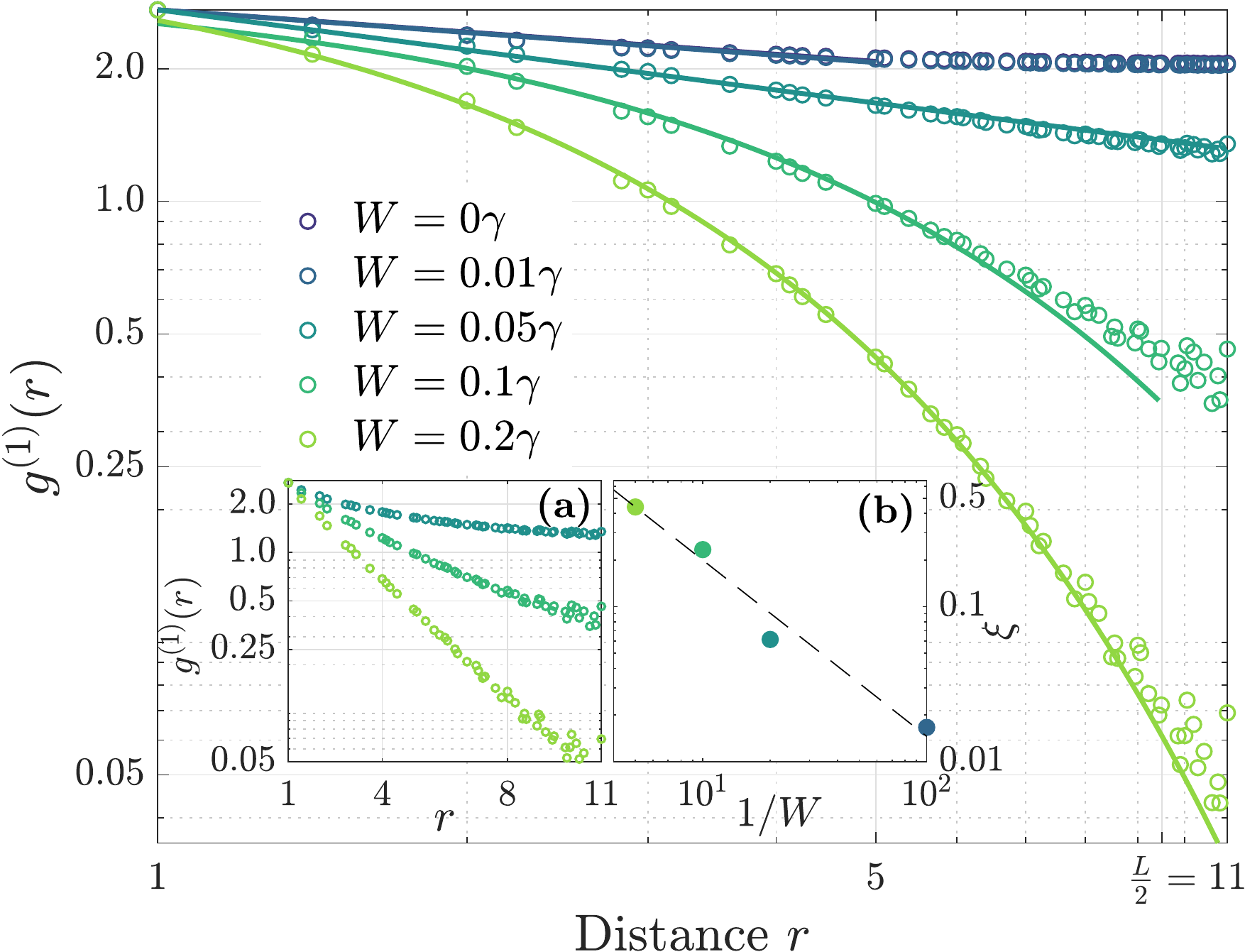}
\caption{Main panel: log-log plot of the two-point correlation function $g^{(1)}(r)$ in a $22\cross 22$ lattice with periodic boundary conditions at $F=1.569\gamma\approx F_{crit}$ for increasing disorder strength $W$. 
The dots are the numerical data while the solid lines are the power-law (for $W=0$) and exponential fits (for $W>0$). 
The other parameters are as in Fig. \ref{fig:pn-disorder}. 
Inset (a): semilog plot of the same data in the main panel.
Inset (b): log-log plot of the extrapolated correlation length $\lambda(W)$ for $W>0$. The dashed line is the fit $\lambda(W)\propto W^{-\eta}$ with $\eta=1.1\pm0.2$.}
\label{fig:Correlations_Disorder}
\end{figure}

Results are shown in Fig. \ref{fig:Correlations_Disorder} \footnote{Each point is obtained considering the mean value of the correlator Eq.\eqref{eq:corrfun} averaged over all the pairs of sites at a distance $r$}. 
In the clean case ($W=0$), the system displays long-range order and the correlation function decays as a power-law 
\begin{equation}
g^{(1)}(r) \sim r^{-\alpha}.
\end{equation}
The exponent $\alpha=0.16 \pm 0.01$ is obtained by fitting the data for $L=22$. 
This result is somehow unexpected since in proximity of a first-order transition one would not expect a divergent correlation length \cite{sachdev2001quantum}.
However, this is a peculiar feature of a perfectly clean system and disorder suppresses long-range order. 

Indeed, in a disordered system ($W\neq0$), the correlation function (\ref{eq:corrfun}) decays exponentially as
\begin{equation}
g^{(1)}(r) \sim {\rm e}^{-r/\lambda(W)},
\end{equation}
where $\lambda(W)\sim W^{-\eta}$ (with $\eta=1.1\pm0.2$) is the correlation length which decreases as $W$ is increased.
This is a consequence of the formation of coherent domains with typical size $\lambda(W)$. 
This means that as long as one considers arrays with $L\lesssim\lambda(W)$, the system will behave cooperatively.
Indeed, for a system of size $L$, we do expect departure from the critical thermodynamic behavior when  $L>\lambda(W)\sim W^{-\eta}$. 
This is consistent to what has been shown in Fig. \ref{fig:scaling_S_W}, since we find that $1/W^*\sim L^{1/\eta}$.
Let us conclude this section remarking that all the exponent obtained here are not expected to be universal since the transition is of first order \cite{Vicentini2018}.

\section{Conclusions and outlook}
\label{sec:conclu}

In this work we have presented a general sampling method to efficiently compute configuration averages of observables in disordered open systems, whose dynamics has been unraveled with stochastic trajectories. 
We have proven that the optimal strategy is realized when each trajectory evolves according to a different disorder configuration. 
Our approach allows us to drastically reduce the computational cost needed to perform disorder-averaged quantities, because the scaling of the statistical error with the number of trajectories and disorder configurations is independent of the specific stochastic evolution protocol and of the particular disorder distribution
Remarkably, the present method can be applied to any Markovian non-equilibrium quantum system numerically solved via stochastic differential equations. 
Our method can represent an invaluable tool to address the dynamical and steady-state properties of a wide class of driven-dissipative systems.  

As a first application, we have been able to study the role of disorder in the driven-dissipative Bose-Hubbard model both in the strongly and weakly interacting regime.
In the case of strong interactions we have analyzed the emergence of spectral resonances due to fermionization of bosons in a disordered 1D chain.
In the weakly interacting case we have explored the role of disorder on a dissipative first-order phase transition occurring in 2D lattices, showing how the criticality is suppressed. 
Thanks to the optimal unraveling method, we have been able to show that the correlation functions of the disordered system decay exponentially in space while for the clean case they decay algebraically at criticality. We have also shown that the derivative of the boson density scales critically (as in the clean system) up to sizes comparable with the typical size of the domains.
The emerging picture is directly relevant for the ongoing experiments in state-of-art photonic quantum simulators based on superconducting circuits and semiconductor microcavities. 
In these systems, the presence of on-site disorder in the local frequency of resonators is unavoidable.
Experimentally, by tuning the size (and/or the disorder) of a sample, one can explore different regimes where collective critical dynamics or density domains dominates the physics.

Many interesting perspective are open for the future. 
The case of spatially correlated noise can be treated within our formalism. 
The combination between our stochastic trajectory approach with cluster techniques \cite{BiellaNLCE2018,Jin16} and tensor-network ansatz \cite{OrusRev2014} represents an intriguing direction. 
Furthermore, developing exact methods to calculate exactly the average Liouvillian dynamics can be useful to shade light on general properties on open many-body disordered systems.
It would be also interesting to study the impact of disorder on a second-order dissipative criticality and to study models where long-range interactions competes with local disorder \cite{Santos2016}.

\begin{acknowledgments}
We thank N. Bartolo, N. Carlon-Zambon, I. Carusotto for stimulating discussions.
We acknowledge support from ERC (via Consolidator Grant CORPHO No.~616233). This work was granted access to the HPC resources of TGCC under the allocations 2018-AP010510493 and 2018-A0050510601 attributed by GENCI ("Grand Equipement National de Calcul Intensif").
\end{acknowledgments}

\bibliography{biblio}
\end{document}